\documentclass[letter]{aa} 

\usepackage{graphicx}
\usepackage{epstopdf}
\usepackage{txfonts}
\usepackage{natbib}
\titlerunning{Rotation of the solar core}

\begin{document}

\title{Rotation rate of the solar core as a key constraint to magnetic angular momentum transport in stellar interiors}

\author{P. Eggenberger\inst{1}
\and G.~Buldgen\inst{1}
\and S.J.A.J.~Salmon\inst{2}}

\institute{Observatoire de Gen\`eve, Universit\'e de Gen\`eve, 51 Ch. des Maillettes, CH-1290 Versoix, Suisse \\  
\email{patrick.eggenberger@unige.ch}
\and
STAR Institute, Universit\'e de Li\`ege, All\'ee du Six Ao\^ut 19C, B-4000 Li\`ege, Belgium
}
   \date{Received; accepted}

 
  \abstract
   {The internal rotation of the Sun constitutes a fundamental constraint when modelling angular momentum transport in stellar interiors. In addition to the more external regions of the solar radiative zone probed by pressure modes, measurements of rotational splittings of gravity modes would offer an invaluable constraint on the rotation of the solar core.}
   {We study the constraints that a measurement of the core rotation rate of the Sun could bring on magnetic angular momentum transport in stellar radiative zones.
}
   {Solar models accounting for angular momentum transport by hydrodynamic and magnetic instabilities were computed for different initial velocities and disc lifetimes on the pre-main sequence to reproduce the surface rotation velocities observed for solar-type stars in open clusters. The internal rotation of these solar models was then compared to helioseismic measurements.}
   {We first show that models computed with angular momentum transport by magnetic instabilities and a recent prescription for the braking of the stellar surface by magnetized winds can reproduce the observations of surface velocities of stars in open clusters. These solar models predict both a flat rotation profile in the external part of the solar radiative zone probed by pressure modes and an increase in the rotation rate in the solar core, where the stabilizing effect of chemical gradients plays a key role. A rapid rotation of the core of the Sun, as suggested by reported detections of gravity modes, is thus found to be compatible with angular momentum transport by magnetic instabilities. Moreover, we show that the efficiency of magnetic angular momentum transport in regions of strong chemical gradients can be calibrated by the solar core rotation rate independently from the unknown rotational history of the Sun. In particular, we find that a recent revised prescription for the transport of angular momentum by the Tayler instability can be easily distinguished from the original Tayler-Spruit dynamo, with a faster rotating solar core supporting the original prescription.
}
{By calibrating the efficiency of magnetic angular momentum transport in regions of strong chemical gradients, a determination of the solar core rotation rate through gravity modes is of prime relevance not only for the Sun, but for stars in general, since radial differential rotation precisely develops in these regions during the more advanced stages of evolution.
}


   \maketitle
%

\section{Introduction}
\label{intro}

The solar five-minute oscillations have led to the determination of the rotation profile of the Sun \citep{kos88,bro89,els95,kos97,cou03,gar07}. This fundamental constraint for the modelling of angular momentum (AM) transport in stellar interiors reveals that rotation is nearly uniform in the solar radiative zone down to about 0.2\,R$_{\odot}$. Interestingly, solar models that account for hydrodynamic transport processes predict a high degree of radial differential rotation in the radiative interior of the Sun, in strong contradiction with helioseismic measurements \citep{pin89,cha95,egg05_mag,tur10}. This indicates that an efficient additional mechanism for the AM transport must be at work in the solar radiative zone.

As recognized long ago, even a weak magnetic field can have a significant impact on the internal rotation of a star \citep[][]{mes53}. Large-scale fossil magnetic fields can be invoked to explain the absence of radial differential rotation in the solar radiative zone \citep[see e.g.][]{mes87,cha93,rue96,spa10}. A key issue is then related to the coupling between the radiative interior and the convective envelope of the Sun. Helioseismic data indicate that there is no differential rotation in the solar radiative zone, while latitudinal differential rotation is present in the convective envelope. If a large-scale magnetic field connects the radiative zone to the convective envelope, then the latitudinal differential rotation of the envelope could be imposed on the radiative interior, in contradiction with helioseismic constraints. This illustrates the difficulty of simultaneously reproducing the nearly uniform rotation in the radiative interior of the Sun with the sharp transition to latitudinal differential rotation in its convective envelope. 

One way to reproduce these helioseismic data is to confine the large-scale field in the solar radiative interior. \cite{gou98} proposed a model that explains how the confinement of the field could be achieved. Numerical simulations have been computed to try to reproduce the configuration proposed by \cite{gou98} and have led to contrasting results \citep[see e.g.][]{bru06,str11,ace13,woo18}. Moreover, a significant increase in the viscosity seems to be required in addition to a large-scale fossil magnetic field in the solar radiative interior to correctly reproduce the low degree of radial differential rotation deduced from helioseismic measurements \citep{rue96,spa10}.
 
It is then particularly interesting to consider the role of magnetic instabilities in the transport of AM. This has been reviewed by \cite{spr99} who found, in stellar radiative zones, that the Tayler instability \citep{tay73} is likely to be the first instability to set in. Based on this instability of the toroidal field and the winding-up of an initial weak field by differential rotation, \cite{spr02} proposed a field amplification cycle that produces a small-scale time-dependent dynamo. Interestingly, the amplitudes of these self-sustained magnetic fields are much larger in the horizontal than in the vertical direction. Consequently, such a small-scale dynamo (known as the Tayler-Spruit dynamo) can correctly account for the sharp transition from uniform rotation in the solar radiative zone to latitudinal differential rotation in the convective envelope. Numerical simulations have been performed to test the viability of this process in stellar radiative zones. Simulations by \cite{bra06} indicate that such a small-scale dynamo can be at work in a differentially rotating stable stratification, while no sign of a dynamo process was found by \cite{zah07}. A definitive answer about the existence of this mechanism would require numerical simulations performed under more realistic stellar conditions, which is of course particularly difficult to achieve \citep{bra17}.

In this context, direct observational constraints on the internal rotation of stars are of prime importance in order to improve the modelling of AM transport in stellar interiors. For the Sun, the inclusion of the Tayler-Spruit dynamo leads to a predicted solar rotation profile in good agreement with the internal rotation deduced from the rotational frequency splittings of pressure (p-) modes \citep{egg05_mag}. Measurements of rotational splittings of gravity (g-) modes would offer invaluable constraints on the rotation of the solar core in addition to the more external regions probed by p-modes. With the increasing total duration of GOLF data, achieving such a detection becomes a real possibility \citep[e.g.][]{bel09,app10,bel11}. As an illustration of this, a recent detection of g-modes has been reported by \cite{fos17}. Similarly to a previous reported detection  \citep{gar07}, it is interesting to note that a fast rotating solar core is favoured by the results of \cite{fos17}.  
Due to the extreme difficulty of detecting g-modes, the robustness of this detection is not guaranteed \citep[see e.g.][]{sch18,app19}. Nevertheless, the possibility of detecting g-modes and probing the rotation of the solar core motivates us to investigate in more detail how the modelling of magnetic AM transport could benefit from such a detection. In particular, is a magnetic transport of AM compatible with both the nearly uniform rotation deduced from p-modes and a fast rotating solar core? More importantly, would a determination of the solar core rotation rate be able to directly constrain the efficiency of AM transport in the core of the Sun without being impacted by its unknown past rotational evolution?   

The input physics and the rotational evolution of models accounting for hydrodynamic and magnetic AM transport processes are described in Sect.~\ref{rotevol}. The internal rotation of these solar models is studied and compared to helioseismic constraints in Sect.~\ref{introt}. The conclusion is given in Sect.~\ref{conclusion}.

\section{Rotational evolution of the models}
\label{rotevol}

Stellar models are computed in the framework of the assumption of shellular rotation \citep{zah92} with the Geneva stellar evolution code \citep{egg08}. The internal transport of angular momentum is followed simultaneously to the evolution of the star by taking into account meridional circulation, shear instability, and magnetic fields in the context of the Tayler-Spruit dynamo \citep[see Sect. 2.1 of][for more details]{egg10_sl}. 

We refer to \cite{mae05} for details about the exact modelling of the Tayler-Spruit dynamo in the Geneva code. We recall here the values of the effective viscosity ($\nu_{\rm TS}$) and the minimum value of radial differential rotation needed for the dynamo process to operate ($q_{\rm min}$) in the two cases originally discussed by \cite{spr02}. When the stratification is dominated by the gradient of chemical composition,
 
\begin{equation}
 \nu_{\rm TS}= r^2 \Omega q^2 \left(\frac{\Omega}{N_{\mu}}\right)^4 \; ; \; q_{\rm min } = \left(\frac{N_{\mu}}{\Omega}\right)^{7/4} \left(\frac{\eta}{r^2 N_{\mu}}\right)^{1/4} \; ,
 \label{nu_gradmu}
 \end{equation}
\noindent while the case of a stratification dominated by the entropy gradient leads to
\begin{equation}
 \nu_{\rm TS} = r^2 \Omega \left(\frac{\Omega}{N_T}\right)^{1/2} \left(\frac{K}{r^2 N_T}\right)^{1/2} \; ; \; q_{\rm min } = \left(\frac{N_T}{\Omega}\right)^{7/4} \left(\frac{\eta}{r^2 N_T}\right)^{1/4} \left(\frac{\eta}{K}\right)^{3/4} \; ,
 \label{nu_NT}
\end{equation}

\noindent with $r$ the radius, $\Omega$ the mean angular velocity, $q= -\frac{\partial \ln \Omega}{\partial \ln r}$, $K$ the thermal diffusivity, $\eta$ the magnetic diffusivity, and $N_T$ and $N_{\mu}$ the thermal and chemical composition part of the Brunt-V\"{a}is\"{a}l\"{a} frequency ($N^2=N_T^2+N_{\mu}^2$).

Including these hydrodynamic and magnetic AM transport processes, the evolution of 1\,M$_{\odot}$ models is computed from the pre-main sequence (PMS) to the solar age. Atomic diffusion is taken into account with diffusion coefficients computed according to the prescription by \cite{paq86}. Braking of the stellar surface by magnetized winds is modelled by adopting the braking law of \cite{mat15,mat19}. The corresponding torque is then given by
\begin{eqnarray}
\nonumber
 \frac{{\rm d} J}{{\rm d}t} = \left\{
\begin{array}{l l }
-T_{\odot} \left({\displaystyle \frac{R}{R_\odot}} \right)^{3.1} 
\left({\displaystyle \frac{M}{M_\odot} }\right)^{0.5} \left({\displaystyle \frac{\tau_{\rm cz}}{\tau_{{\rm cz}\odot}} }\right)^{p} \left({\displaystyle \frac{\Omega}{\Omega_{\odot}} }\right)^{p+1} & 
(Ro > Ro_{\odot}/\chi) \\
\nonumber
 & \\
\nonumber
-T_{\odot} \left({\displaystyle \frac{R}{R_\odot}} \right)^{3.1} 
\left({\displaystyle \frac{M}{M_\odot} }\right)^{0.5} \chi^{p} \left({\displaystyle \frac{\Omega}{\Omega_{\odot}} }\right)
& (Ro \leq Ro_{\odot}/\chi) \; ,  
 \end{array}   \right.
\end{eqnarray}

\noindent with $R$ and $M$ the stellar radius and mass, $Ro$ the Rossby number, and $\tau_{\rm cz}$ the convective turnover timescale. The constant $\chi$ defines the transition from saturated to unsaturated regime ($\chi=Ro_{\odot}/Ro_{\rm sat}=\Omega_{\rm sat}\tau_{\rm cz}/\Omega_{\odot}\tau_{{\rm cz}\odot}$) and is fixed to 10. The coefficient $p$ is taken equal to 2.3 and the braking constant $T_{\odot}$ is calibrated to reproduce the surface rotation rate of the Sun. 
In convective zones, a very efficient transport of AM is assumed, resulting in a flat rotation profile in these zones. The initial chemical composition, the mixing-length parameter for convection, and the braking constant related to magnetized winds are then calibrated to reproduce the solar photospheric abundances as given by \cite{gre93}, the solar luminosity, radius, and surface rotational velocity after $4.57$\,Gyr.

\begin{figure}[htb!]
\resizebox{\hsize}{!}{\includegraphics{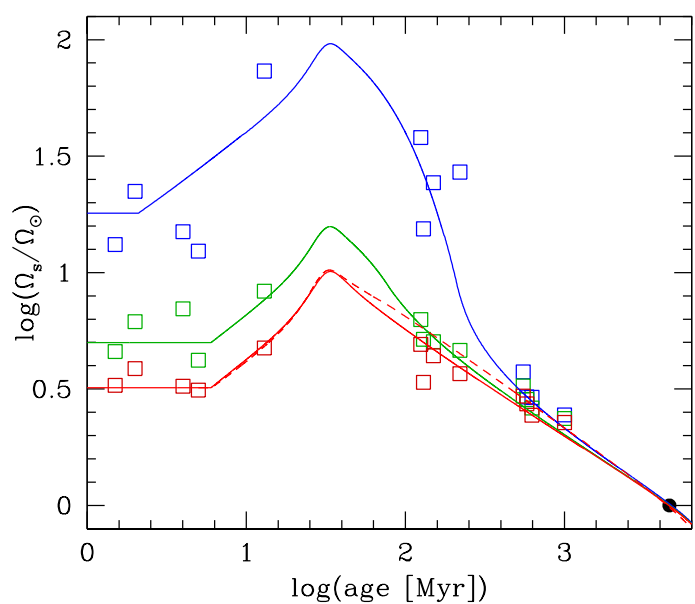}}
 \caption{Surface angular velocity $\Omega_{\rm s}$ as a function of age for solar models. Blue, green, and red lines indicate models of fast, moderate, and slow rotators, respectively. Open symbols show observations of surface rotation rates in open clusters taken from Table 1 of \cite{gal15}, with blue, green, and red symbols indicating the 90th, 50th, and 25th rotational percentiles, respectively. The black dot indicates the surface rotation of the Sun. Continuous lines refer to models computed with the surface braking law of \cite{mat15}, while the dashed red line corresponds to a model computed with the braking law of \cite{kri97}.}
  \label{omegas}
\end{figure}

The rotational history of the Sun being unknown, we consider different models representative of slow, moderate, and fast rotators as deduced from observations of surface rotation rates of solar-type stars in open clusters of various ages. The initial rotation rates and disc lifetimes during the PMS are varied in order to try to reproduce the 90th, 50th, and 25th rotational percentiles (corresponding to fast, moderate, and slow rotating cases, respectively) in these clusters, as given by \cite{gal15}. The evolution of the surface angular velocity of these models as a function of time is shown in Fig.~\ref{omegas}. The fast rotating model (blue line) is computed with an initial angular velocity of 18\,$\Omega_\odot$ and a disc lifetime of 2\,Myr. The moderate (green line) and slow (red line) rotating models share the same disc lifetime of 6\,Myr and have an initial velocity of 5\,$\Omega_\odot$ and 3.2\,$\Omega_\odot$, respectively. During the disc-locking phase, we simply assume that the surface angular velocity of the star remains constant. After this phase, the surface velocity rapidly increases due to the PMS contraction. The zero age main sequence (ZAMS) is reached at an age of about 40\,Myr with a surface velocity close to its maximum value. Magnetized winds are then responsible for the decrease in the surface rotation of the models during the main sequence (MS) and to the convergence of their surface rotation rates to reach the solar value after $4.57$\,Gyr.

Figure~\ref{omegas} shows that rotating models computed with the Tayler-Spruit dynamo and the braking law of \cite{mat15} correctly reproduce the surface rotation rates observed for solar-type stars in open clusters. This is an interesting result, because a previous study by \cite{den10_spin} suggested that the efficient transport of AM induced by the Tayler-Spruit dynamo is not compatible with the surface rotation observed for slowly rotating stars. The study by \cite{den10_spin} used the braking law of \cite{kri97}, while our results are obtained with the more recent prescription of \cite{mat15}. To investigate the influence of the adopted braking law on these conclusions, an additional model is computed for the slow rotating case using the \cite{kri97} law as in \cite{den10_spin}. The evolution of the surface velocity of this model is shown by the dashed red line in Fig.~\ref{omegas}. Adopting the braking law of \cite{kri97} effectively leads to  a surface rotation rate that is too high during the beginning of the MS for slowly rotating stars, in good agreement with the results of \cite{den10_spin}. This illustrates the sensitivity of the predicted surface rotation rates on the uncertainties related to the adopted braking law.

\section{Internal rotation of the solar models}
\label{introt}

\begin{figure}[htb!]
\resizebox{\hsize}{!}{\includegraphics{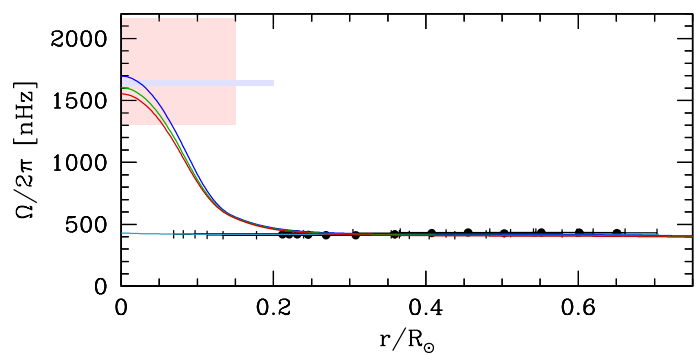}}
 \caption{Rotation profiles for solar models computed with different initial rotation velocities and disc lifetimes. Blue, green, and red continuous lines correspond to the fast, moderate, and slow rotating models shown in Fig.~\ref{omegas}. The light blue line indicates the model computed with the revised prescription of \cite{ful19}. Black dots correspond to the internal rotation in the solar radiative zone as deduced from p-modes \citep{cou03}. The red and blue regions indicate the mean rotation rate in the solar core corresponding to the detections of g-modes reported by \cite{gar07} and \cite{fos17}, respectively.}
  \label{pro_omega_qmin}
\end{figure}

Figure~\ref{pro_omega_qmin} shows the rotation profile of the solar models corresponding to the fast (blue), moderate (green), and slow (red) rotating cases discussed above. We note that these models exhibit higher core rotation rates than previous models computed by \cite{egg05_mag} due to the different input physics used. In particular, the present models correspond to complete solar models computed with atomic diffusion, rotation, and magnetic fields from the PMS to the solar age, while \cite{egg05_mag} simply focussed on the evolution of the internal rotation of 1\,M$_{\odot}$ models computed without atomic diffusion and assuming solid-body rotation on the ZAMS. All models correctly reproduce the rotation rates in the external part of the radiative zone (above 0.2\,R$_{\odot}$) as probed by p-modes (black dots in Fig.~\ref{pro_omega_qmin}). A key feature of these models is that they predict an increase in the rotation rate in the solar core (below about 0.2\,R$_{\odot}$). We thus find that a rapid rotation of the core of the Sun can be reproduced by an efficient transport of AM by magnetic instabilities, which is required to explain the nearly uniform rotation of the solar radiative zone deduced from helioseismic measurements of p-modes. This is an important result regarding the possibility of deducing the solar core rotation rate through the detection of rotational splittings of g-modes. In particular, reported detections of g-modes suggest that the core of the Sun could be rotating more rapidly than the rest of the radiative zone \citep[][]{gar07,gar11,fos17}.

\begin{figure}[htb!]
\resizebox{\hsize}{!}{\includegraphics{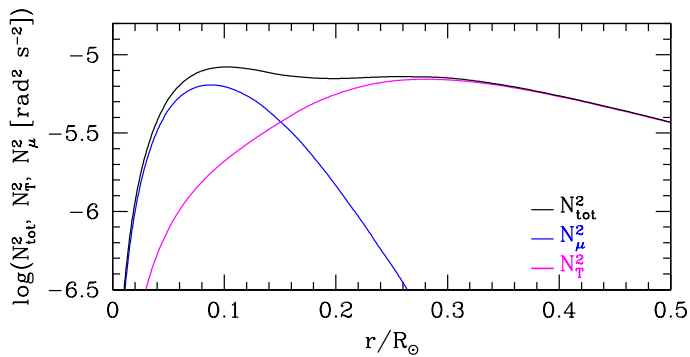}}
 \caption{Brunt-V\"{a}is\"{a}l\"{a} frequencies $N_{\rm tot}^2$, $N_{\mu}^2$, and $N_T^2$ ($N_{\rm tot}^2=N_{\mu}^2 + N_T^2$) in the central layers of the solar model computed in the case of a moderate initial rotation rate.}
  \label{evogradmu}
\end{figure}

The change in rotation rates between the core and the more external part of the solar radiative zone is related to the stabilizing role of the gradients of chemical composition. In the solar interior, stratification is due to both chemical and thermal gradients. In the case of hydrodynamic and magnetic instabilities, thermal diffusion efficiently reduces the stabilizing effect of the entropy gradient \cite[e.g.][]{zah74,spr99}; however, thermal diffusion does not reduce the strong inhibiting effect of the chemical gradients. These gradients of chemical composition develop in the central parts of the star during its MS evolution due to nuclear reactions. This is illustrated in Fig.~\ref{evogradmu}, which shows the chemical ($N_{\mu}^2$) and thermal ($N_T^2$) parts of the Brunt-V\"{a}is\"{a}l\"{a} frequency for the solar model computed in the moderate rotating case. In the Sun, the contribution of chemical gradients to the total Brunt-V\"{a}is\"{a}l\"{a} frequency becomes significant below 0.3\,R$_{\odot}$, and $N_{\mu}^2$ is dominating over $N_T^2$ below about 0.15\,R$_{\odot}$.   

The impact of chemical gradients on the solar core rotation rate can be understood by comparing the equations of the Tayler-Spruit dynamo in the cases where the stratification is dominated by the chemical (Eq.~\ref{nu_gradmu} above) and thermal (Eq. \ref{nu_NT}) gradients. In Eq.~\ref{nu_gradmu} we first note that the efficiency associated with the transport of AM by the Tayler-Spruit dynamo is strongly reduced when the gradient of molecular weight increases ($\nu_{\rm TS} \propto N_{\mu}^{-4} \propto \nabla_{\mu}^{-2}$). In addition to the decrease in the AM transport efficiency, chemical gradients also have an important effect on the minimum value of radial differential rotation needed for the dynamo to operate ($q_{\rm min}$). Due to the reduction of the stabilizing effect of the entropy gradient by thermal diffusion, the expression of $q_{\rm min}$ in regions where chemical gradients dominate (Eq.~\ref{nu_gradmu}) is indeed the same as that obtained when thermal stratification dominates (Eq.~\ref{nu_NT}) given that $N_{\mu}^2$ is replaced by an effective frequency $N_{T,{\rm eff}}^2 \approx \frac{\eta}{K} N_T^2$ \cite[][]{spr99}. The ratio of magnetic to thermal diffusivity being much lower than one, the minimum value of radial differential rotation needed for the dynamo to operate is much higher in regions with strong chemical gradients. The decrease in the viscosity and the increase in the value of the minimum differential rotation $q_{\rm min}$ induced by chemical gradients both participate in the rapid transition from a nearly flat rotation profile above 0.2\,R$_{\odot}$ to a faster rotating core seen in Fig.~\ref{pro_omega_qmin}.

The increase in the rotation rate in the solar core is very similar for models computed for different initial rotational velocities. The core rotation rate of the Sun as predicted by models accounting for AM transport by magnetic instabilities is thus found to be almost insensitive to its rotational history. This is a fundamental result, since it implies that a measurement of the solar core rotation rate would directly constrain the efficiency of AM transport by magnetic instabilities without being impacted by the unknown rotational history of the Sun. This is of particular interest because current expressions for the Tayler-Spruit dynamo were obtained by neglecting multiplying factors of order unity \cite[see conclusion of][]{spr02}. 
Moreover, a revised prescription for AM transport by the Tayler instability has recently been proposed \citep{ful19}. The effective viscosity and minimum radial differential rotation needed for this process to operate are given by \citep[Eqs. 35 and 36 in][]{ful19}
\begin{equation}
\nu_{\rm T}= \alpha^{3} r^2 \Omega \left(\frac{\Omega}{N_{\rm eff}}\right)^2  \; ; \; q_{\rm min } = \alpha^{-3}\left(\frac{N_{\rm eff}}{\Omega}\right)^{5/2} \left(\frac{\eta}{r^2 \Omega}\right)^{3/4} \; ,
 \label{nu_F}
 \end{equation} 
with $\alpha$ a dimensionless parameter fixed to 1 and $N_{\rm eff}$ the effective Brunt-V\"{a}is\"{a}l\"{a} frequency given by $N^2_{\rm eff}= \frac{\eta}{K} N_T^2 + N_{\mu}^2$. An additional model is then computed for the moderate rotating case using Eq.~(\ref{nu_F}) proposed by \cite{ful19} instead of the original Tayler-Spruit dynamo. The effective viscosity $\nu_{\rm T}$ in the solar radiative zone associated with the AM transport by the Tayler instability as given by Eq.~(\ref{nu_F}) is typically 4 orders of magnitude higher than that associated with the Tayler-Spruit dynamo. Owing to this very efficient AM transport, the prescription recently proposed by \cite{ful19} predicts an almost uniform rotation down to the solar centre (light blue line in Fig.~\ref{pro_omega_qmin}). Models computed with the original Tayler-Spruit dynamo and with the prescription of \cite{ful19} can then be easily distinguished by a determination of the solar core rotation rate. As shown in Fig.~\ref{pro_omega_qmin} a fast rotating solar core, as suggested by reported detections of g-modes, favours the original prescription for the Tayler-Spruit dynamo\footnote{Even these models predict a slightly lower mean rotation rate of the solar radiative zone than the value reported by \cite{fos17}.}.

\section{Conclusion}
\label{conclusion}

Solar models that account for AM transport by meridional circulation, shear instability, and magnetic instabilities were first computed for different initial velocities and disc lifetimes on the PMS. Using the prescription of \cite{mat15} for the braking of the stellar surface by magnetized winds, we find that the evolution of the surface velocity of these models can reproduce the surface rotation rates observed for solar-type stars in open clusters. 

We then study the internal rotation of these solar models. An almost uniform rotation is predicted above 0.2\,R$_{\odot}$, in good agreement with helioseismic measurements of p-modes, while an increase in the rotation rate is found in the solar core. The change in rotation rates between the core and the more external part of the solar radiative zone seen in these models is a key signature of the inhibiting effect of chemical gradients on the transport of AM by magnetic instabilities. We thus find that a solar core rotating faster than the more external part of the radiative zone, as suggested by reported detections of g-modes, can be correctly reproduced by models accounting for AM transport by magnetic instabilities. Should the detection of a fast rotating solar core be confirmed, this would strongly support magnetic instabilities as the main AM transport process in the solar radiative zone. In addition to magnetic AM transport, it would be interesting to perform the same kind of comparison with solar models that account for AM transport by internal gravity waves \citep[e.g.][]{zah97,cha05,pin17}.

We also show that the efficiency of AM transport by magnetic instabilities in regions of strong chemical gradients can be directly calibrated by the measurement of the solar core rotation rate independently from the unknown past rotational evolution of the Sun. We then find that the revised prescription for AM transport by the Tayler instability recently proposed by \cite{ful19} can be easily distinguished from the original prescription of the Tayler-Spruit dynamo, with a fast rotating solar core in favour of the original prescription proposed by \cite{spr02}. 

We conclude that the core rotation rate of the Sun constitutes the perfect constraint to calibrate the efficiency of magnetic AM transport in regions of strong chemical gradients. This is of prime interest not only for the specific case of solar models, but for stellar evolution in general, since radial differential rotation precisely develops in layers with strong gradients of chemical composition \citep[e.g.][]{heg05,mae05}. Calibrating the AM transport efficiency in these regions from a firm detection of solar g-modes would ideally complement constraints on AM transport that are currently being obtained from comparisons between rotating models and rotational splittings of mixed modes in subgiants \cite[e.g.][]{deh14,pin17,egg19} and red giant stars \cite[e.g.][]{egg12_rg,mar13,cei13,can14,spa16,egg17}.

\begin{acknowledgements}
We thank the referee for the useful comments that helped us improve the quality of the manuscript.
This work has been supported by the Swiss National Science Foundation (project Interacting Stars, number 200020-172505).
\end{acknowledgements}


\bibliographystyle{aa} 
\bibliography{biblio} 

\begin{thebibliography}{58}
\expandafter\ifx\csname natexlab\endcsname\relax\def\natexlab#1{#1}\fi

\bibitem[{{Acevedo-Arreguin} {et~al.}(2013){Acevedo-Arreguin}, {Garaud}, \&
  {Wood}}]{ace13}
{Acevedo-Arreguin}, L.~A., {Garaud}, P., \& {Wood}, T.~S. 2013, \mnras, 434,
  720

\bibitem[{{Appourchaux} {et~al.}(2010){Appourchaux}, {Belkacem}, {Broomhall},
  {Chaplin}, {Gough}, {Houdek}, {Provost}, {Baudin}, {Boumier}, {Elsworth},
  {Garc{\'{\i}}a}, {Andersen}, {Finsterle}, {Fr{\"o}hlich}, {Gabriel}, {Grec},
  {Jim{\'e}nez}, {Kosovichev}, {Sekii}, {Toutain}, \&
  {Turck-Chi{\`e}ze}}]{app10}
{Appourchaux}, T., {Belkacem}, K., {Broomhall}, A.-M., {et~al.} 2010, \aapr,
  18, 197

\bibitem[{{Appourchaux} \& {Corbard}(2019)}]{app19}
{Appourchaux}, T. \& {Corbard}, T. 2019, \aap, 624, A106

\bibitem[{{Belkacem}(2011)}]{bel11}
{Belkacem}, K. 2011, in Lecture Notes in Physics, Berlin Springer Verlag, Vol.
  832, Lecture Notes in Physics, Berlin Springer Verlag, ed. J.-P. {Rozelot} \&
  C.~{Neiner}, 139

\bibitem[{{Belkacem} {et~al.}(2009){Belkacem}, {Samadi}, {Goupil}, {Dupret},
  {Brun}, \& {Baudin}}]{bel09}
{Belkacem}, K., {Samadi}, R., {Goupil}, M.~J., {et~al.} 2009, \aap, 494, 191

\bibitem[{{Braithwaite}(2006)}]{bra06}
{Braithwaite}, J. 2006, \aap, 449, 451

\bibitem[{{Braithwaite} \& {Spruit}(2017)}]{bra17}
{Braithwaite}, J. \& {Spruit}, H.~C. 2017, Royal Society Open Science, 4,
  160271

\bibitem[{{Brown} {et~al.}(1989){Brown}, {Christensen-Dalsgaard},
  {Dziembowski}, {Goode}, {Gough}, \& {Morrow}}]{bro89}
{Brown}, T.~M., {Christensen-Dalsgaard}, J., {Dziembowski}, W.~A., {et~al.}
  1989, \apj, 343, 526

\bibitem[{{Brun} \& {Zahn}(2006)}]{bru06}
{Brun}, A.~S. \& {Zahn}, J.-P. 2006, \aap, 457, 665

\bibitem[{{Cantiello} {et~al.}(2014){Cantiello}, {Mankovich}, {Bildsten},
  {Christensen-Dalsgaard}, \& {Paxton}}]{can14}
{Cantiello}, M., {Mankovich}, C., {Bildsten}, L., {Christensen-Dalsgaard}, J.,
  \& {Paxton}, B. 2014, \apj, 788, 93

\bibitem[{{Ceillier} {et~al.}(2013){Ceillier}, {Eggenberger}, {Garc{\'{\i}}a},
  \& {Mathis}}]{cei13}
{Ceillier}, T., {Eggenberger}, P., {Garc{\'{\i}}a}, R.~A., \& {Mathis}, S.
  2013, \aap, 555, A54

\bibitem[{{Chaboyer} {et~al.}(1995){Chaboyer}, {Demarque}, \&
  {Pinsonneault}}]{cha95}
{Chaboyer}, B., {Demarque}, P., \& {Pinsonneault}, M.~H. 1995, \apj, 441, 865

\bibitem[{{Charbonneau} \& {MacGregor}(1993)}]{cha93}
{Charbonneau}, P. \& {MacGregor}, K.~B. 1993, \apj, 417, 762

\bibitem[{{Charbonnel} \& {Talon}(2005)}]{cha05}
{Charbonnel}, C. \& {Talon}, S. 2005, Science, 309, 2189

\bibitem[{{Couvidat} {et~al.}(2003){Couvidat}, {Garc{\'{\i}}a},
  {Turck-Chi{\`e}ze}, {Corbard}, {Henney}, \& {Jim{\'e}nez-Reyes}}]{cou03}
{Couvidat}, S., {Garc{\'{\i}}a}, R.~A., {Turck-Chi{\`e}ze}, S., {et~al.} 2003,
  \apjl, 597, L77

\bibitem[{{Deheuvels} {et~al.}(2014){Deheuvels}, {Do{\u g}an}, {Goupil},
  {Appourchaux}, {Benomar}, {Bruntt}, {Campante}, {Casagrande}, {Ceillier},
  {Davies}, {De Cat}, {Fu}, {Garc{\'{\i}}a}, {Lobel}, {Mosser}, {Reese},
  {Regulo}, {Schou}, {Stahn}, {Thygesen}, {Yang}, {Chaplin},
  {Christensen-Dalsgaard}, {Eggenberger}, {Gizon}, {Mathis},
  {Molenda-{\.Z}akowicz}, \& {Pinsonneault}}]{deh14}
{Deheuvels}, S., {Do{\u g}an}, G., {Goupil}, M.~J., {et~al.} 2014, \aap, 564,
  A27

\bibitem[{{Denissenkov} {et~al.}(2010){Denissenkov}, {Pinsonneault},
  {Terndrup}, \& {Newsham}}]{den10_spin}
{Denissenkov}, P.~A., {Pinsonneault}, M., {Terndrup}, D.~M., \& {Newsham}, G.
  2010, \apj, 716, 1269

\bibitem[{{Eggenberger} {et~al.}(2019){Eggenberger}, {Deheuvels}, {Miglio},
  {Ekstr{\"o}m}, {Georgy}, {Meynet}, {Lagarde}, {Salmon}, {Buldgen},
  {Montalb{\'a}n}, {Spada}, \& {Ballot}}]{egg19}
{Eggenberger}, P., {Deheuvels}, S., {Miglio}, A., {et~al.} 2019, \aap, 621, A66

\bibitem[{{Eggenberger} {et~al.}(2017){Eggenberger}, {Lagarde}, {Miglio},
  {Montalb{\'a}n}, {Ekstr{\"o}m}, {Georgy}, {Meynet}, {Salmon}, {Ceillier},
  {Garc{\'{\i}}a}, {Mathis}, {Deheuvels}, {Maeder}, {den Hartogh}, \&
  {Hirschi}}]{egg17}
{Eggenberger}, P., {Lagarde}, N., {Miglio}, A., {et~al.} 2017, \aap, 599, A18

\bibitem[{{Eggenberger} {et~al.}(2005){Eggenberger}, {Maeder}, \&
  {Meynet}}]{egg05_mag}
{Eggenberger}, P., {Maeder}, A., \& {Meynet}, G. 2005, \aap, 440, L9

\bibitem[{{Eggenberger} {et~al.}(2008){Eggenberger}, {Meynet}, {Maeder},
  {Hirschi}, {Charbonnel}, {Talon}, \& {Ekstr{\"o}m}}]{egg08}
{Eggenberger}, P., {Meynet}, G., {Maeder}, A., {et~al.} 2008, \apss, 316, 43

\bibitem[{{Eggenberger} {et~al.}(2010){Eggenberger}, {Meynet}, {Maeder},
  {Miglio}, {Montalban}, {Carrier}, {Mathis}, {Charbonnel}, \&
  {Talon}}]{egg10_sl}
{Eggenberger}, P., {Meynet}, G., {Maeder}, A., {et~al.} 2010, \aap, 519, A116

\bibitem[{{Eggenberger} {et~al.}(2012){Eggenberger}, {Montalb{\'a}n}, \&
  {Miglio}}]{egg12_rg}
{Eggenberger}, P., {Montalb{\'a}n}, J., \& {Miglio}, A. 2012, \aap, 544, L4

\bibitem[{{Elsworth} {et~al.}(1995){Elsworth}, {Howe}, {Isaak}, {McLeod},
  {Miller}, {New}, {Wheeler}, \& {Gough}}]{els95}
{Elsworth}, Y., {Howe}, R., {Isaak}, G.~R., {et~al.} 1995, \nat, 376, 669

\bibitem[{{Fossat} {et~al.}(2017){Fossat}, {Boumier}, {Corbard}, {Provost},
  {Salabert}, {Schmider}, {Gabriel}, {Grec}, {Renaud}, {Robillot},
  {Roca-Cort{\'e}s}, {Turck-Chi{\`e}ze}, {Ulrich}, \& {Lazrek}}]{fos17}
{Fossat}, E., {Boumier}, P., {Corbard}, T., {et~al.} 2017, \aap, 604, A40

\bibitem[{{Fuller} {et~al.}(2019){Fuller}, {Piro}, \& {Jermyn}}]{ful19}
{Fuller}, J., {Piro}, A.~L., \& {Jermyn}, A.~S. 2019, \mnras, 485, 3661

\bibitem[{{Gallet} \& {Bouvier}(2015)}]{gal15}
{Gallet}, F. \& {Bouvier}, J. 2015, \aap, 577, A98

\bibitem[{{Garc{\'{\i}}a} {et~al.}(2011){Garc{\'{\i}}a}, {Salabert}, {Ballot},
  {Eff-Darwich}, {Garrido}, {Jim{\'e}nez}, {Mathis}, {Mathur}, {Moya},
  {Pall{\'e}}, {R{\'e}gulo}, {Sato}, {Su{\'a}rez}, \&
  {Turck-Chi{\`e}ze}}]{gar11}
{Garc{\'{\i}}a}, R.~A., {Salabert}, D., {Ballot}, J., {et~al.} 2011, GONG-SoHO
  24: A New Era of Seismology of the Sun and Solar-Like Stars, 271, 012046

\bibitem[{{Garc{\'{\i}}a} {et~al.}(2007){Garc{\'{\i}}a}, {Turck-Chi{\`e}ze},
  {Jim{\'e}nez-Reyes}, {Ballot}, {Pall{\'e}}, {Eff-Darwich}, {Mathur}, \&
  {Provost}}]{gar07}
{Garc{\'{\i}}a}, R.~A., {Turck-Chi{\`e}ze}, S., {Jim{\'e}nez-Reyes}, S.~J.,
  {et~al.} 2007, Science, 316, 1591

\bibitem[{{Gough} \& {McIntyre}(1998)}]{gou98}
{Gough}, D.~O. \& {McIntyre}, M.~E. 1998, \nat, 394, 755

\bibitem[{{Grevesse} \& {Noels}(1993)}]{gre93}
{Grevesse}, N. \& {Noels}, A. 1993, in Origin and evolution of the elements:
  proceedings of a symposium in honour of H. Reeves, held in Paris, June 22-25,
  1992. Edited by N. Prantzos, E. Vangioni-Flam and M. Casse. Published by
  Cambridge University Press, Cambridge, England, 1993, p.14, ed.
  N.~{Prantzos}, E.~{Vangioni-Flam}, \& M.~{Casse}, 14

\bibitem[{{Heger} {et~al.}(2005){Heger}, {Woosley}, \& {Spruit}}]{heg05}
{Heger}, A., {Woosley}, S.~E., \& {Spruit}, H.~C. 2005, \apj, 626, 350

\bibitem[{{Kosovichev}(1988)}]{kos88}
{Kosovichev}, A.~G. 1988, Soviet Astronomy Letters, 14, 145

\bibitem[{{Kosovichev} {et~al.}(1997){Kosovichev}, {Schou}, {Scherrer},
  {Bogart}, {Bush}, {Hoeksema}, {Aloise}, {Bacon}, {Burnette}, {de Forest},
  {Giles}, {Leibrand}, {Nigam}, {Rubin}, {Scott}, {Williams}, {Basu},
  {Christensen-Dalsgaard}, {D\"appen}, {Rhodes}, {Duvall}, {Howe}, {Thompson},
  {Gough}, {Sekii}, {Toomre}, {Tarbell}, {Title}, {Mathur}, {Morrison}, {Saba},
  {Wolfson}, {Zayer}, \& {Milford}}]{kos97}
{Kosovichev}, A.~G., {Schou}, J., {Scherrer}, P.~H., {et~al.} 1997, \solphys,
  170, 43

\bibitem[{{Krishnamurthi} {et~al.}(1997){Krishnamurthi}, {Pinsonneault},
  {Barnes}, \& {Sofia}}]{kri97}
{Krishnamurthi}, A., {Pinsonneault}, M.~H., {Barnes}, S., \& {Sofia}, S. 1997,
  \apj, 480, 303

\bibitem[{{Maeder} \& {Meynet}(2005)}]{mae05}
{Maeder}, A. \& {Meynet}, G. 2005, \aap, 440, 1041

\bibitem[{{Marques} {et~al.}(2013){Marques}, {Goupil}, {Lebreton}, {Talon},
  {Palacios}, {Belkacem}, {Ouazzani}, {Mosser}, {Moya}, {Morel}, {Pichon},
  {Mathis}, {Zahn}, {Turck-Chi{\`e}ze}, \& {Nghiem}}]{mar13}
{Marques}, J.~P., {Goupil}, M.~J., {Lebreton}, Y., {et~al.} 2013, \aap, 549,
  A74

\bibitem[{{Matt} {et~al.}(2015){Matt}, {Brun}, {Baraffe}, {Bouvier}, \&
  {Chabrier}}]{mat15}
{Matt}, S.~P., {Brun}, A.~S., {Baraffe}, I., {Bouvier}, J., \& {Chabrier}, G.
  2015, \apjl, 799, L23

\bibitem[{{Matt} {et~al.}(2019){Matt}, {Brun}, {Baraffe}, {Bouvier}, \&
  {Chabrier}}]{mat19}
{Matt}, S.~P., {Brun}, A.~S., {Baraffe}, I., {Bouvier}, J., \& {Chabrier}, G.
  2019, \apjl, 870, L27

\bibitem[{{Mestel}(1953)}]{mes53}
{Mestel}, L. 1953, \mnras, 113, 716

\bibitem[{{Mestel} \& {Weiss}(1987)}]{mes87}
{Mestel}, L. \& {Weiss}, N.~O. 1987, \mnras, 226, 123

\bibitem[{{Paquette} {et~al.}(1986){Paquette}, {Pelletier}, {Fontaine}, \&
  {Michaud}}]{paq86}
{Paquette}, C., {Pelletier}, C., {Fontaine}, G., \& {Michaud}, G. 1986, \apjs,
  61, 177

\bibitem[{{Pin{\c c}on} {et~al.}(2017){Pin{\c c}on}, {Belkacem}, {Goupil}, \&
  {Marques}}]{pin17}
{Pin{\c c}on}, C., {Belkacem}, K., {Goupil}, M.~J., \& {Marques}, J.~P. 2017,
  \aap, 605, A31

\bibitem[{{Pinsonneault} {et~al.}(1989){Pinsonneault}, {Kawaler}, {Sofia}, \&
  {Demarque}}]{pin89}
{Pinsonneault}, M.~H., {Kawaler}, S.~D., {Sofia}, S., \& {Demarque}, P. 1989,
  \apj, 338, 424

\bibitem[{{R\"udiger} \& {Kitchatinov}(1996)}]{rue96}
{R\"udiger}, G. \& {Kitchatinov}, L.~L. 1996, \apj, 466, 1078

\bibitem[{{Schunker} {et~al.}(2018){Schunker}, {Schou}, {Gaulme}, \&
  {Gizon}}]{sch18}
{Schunker}, H., {Schou}, J., {Gaulme}, P., \& {Gizon}, L. 2018, \solphys, 293,
  95

\bibitem[{{Spada} {et~al.}(2016){Spada}, {Gellert}, {Arlt}, \&
  {Deheuvels}}]{spa16}
{Spada}, F., {Gellert}, M., {Arlt}, R., \& {Deheuvels}, S. 2016, \aap, 589, A23

\bibitem[{{Spada} {et~al.}(2010){Spada}, {Lanzafame}, \& {Lanza}}]{spa10}
{Spada}, F., {Lanzafame}, A.~C., \& {Lanza}, A.~F. 2010, \mnras, 404, 641

\bibitem[{{Spruit}(1999)}]{spr99}
{Spruit}, H.~C. 1999, \aap, 349, 189

\bibitem[{{Spruit}(2002)}]{spr02}
{Spruit}, H.~C. 2002, \aap, 381, 923

\bibitem[{{Strugarek} {et~al.}(2011){Strugarek}, {Brun}, \& {Zahn}}]{str11}
{Strugarek}, A., {Brun}, A.~S., \& {Zahn}, J.-P. 2011, \aap, 532, A34

\bibitem[{{Tayler}(1973)}]{tay73}
{Tayler}, R.~J. 1973, \mnras, 161, 365

\bibitem[{{Turck-Chi{\`e}ze} {et~al.}(2010){Turck-Chi{\`e}ze}, {Palacios},
  {Marques}, \& {Nghiem}}]{tur10}
{Turck-Chi{\`e}ze}, S., {Palacios}, A., {Marques}, J.~P., \& {Nghiem}, P.~A.~P.
  2010, \apj, 715, 1539

\bibitem[{{Wood} \& {Brummell}(2018)}]{woo18}
{Wood}, T.~S. \& {Brummell}, N.~H. 2018, \apj, 853, 97

\bibitem[{{Zahn} {et~al.}(2007){Zahn}, {Brun}, \& {Mathis}}]{zah07}
{Zahn}, J., {Brun}, A.~S., \& {Mathis}, S. 2007, \aap, 474, 145

\bibitem[{{Zahn}(1974)}]{zah74}
{Zahn}, J.-P. 1974, in IAU Symposium, Vol.~59, Stellar Instability and
  Evolution, ed. P.~{Ledoux}, A.~{Noels}, \& A.~W. {Rodgers}, 185--194

\bibitem[{{Zahn}(1992)}]{zah92}
{Zahn}, J.-P. 1992, \aap, 265, 115

\bibitem[{{Zahn} {et~al.}(1997){Zahn}, {Talon}, \& {Matias}}]{zah97}
{Zahn}, J.-P., {Talon}, S., \& {Matias}, J. 1997, \aap, 322, 320

\end{thebibliography}

\end{document}